\documentclass[twocolumn,amssymb,amsmath,superscriptaddress]{revtex4}
\usepackage[colorlinks,hypertex,hidelinks]{hyperref}
\usepackage[nice]{nicefrac}
\hypersetup{
    pdfborder={0 0 0},
    colorlinks = true,
    linkcolor = blue,
    anchorcolor = blue,
    citecolor = blue,
    filecolor = blue,
    urlcolor = blue
    }
\usepackage{graphicx}  
\usepackage{bm}
\usepackage{xfrac}
\usepackage{enumitem}
\newcommand{\R}{\text{Re}}

\begin{document}

\title{Chiral Charge Density Waves in Transition Metal Dichalcogenide}
\author{Ettore~Carpene}
\email{ettore.carpene@cnr.it}
\affiliation{CNR-IFN, Institute for Photonics and Nanotechnologies, piazza Leonardo da Vinci 32, 20133 Milano, Italy}
%


\begin{abstract}
The emergence of chirality in the charge density wave (CDW) phase of 1$T$-TiSe$_2$ has recently attracted significant attention, yet its microscopic origin remains debated. The prevailing interpretation attributes chirality to a relative phase shift between the charge density components. Here, using density functional theory and symmetry analysis, it is demonstrated that such phase-shifted states entail prohibitive energy costs, rendering them unlikely candidates for the ground state. Instead, a nearly degenerate metastable CDW phase stacking order with monoclinic symmetry is identified, that naturally breaks mirror and inversion symmetries. The CDW in 1$T$-TiSe$_2$ possesses a dual physical nature: the electronic charge redistribution is intrinsically bond-modulated, while the periodic lattice distortion is atom-centered transverse. This spatial separation allows a specific optical coupling mechanism where circularly polarized light breaks the energetic degeneracy between chiral domains. These results reconcile conflicting experimental observations, identifying the "spontaneous" chirality observed in local probes as a statistical distribution of CDW phase stacking domains, while establishing the mechanism for macroscopic, light-induced gyrotropic order.
\end{abstract}
\maketitle

Transition metal dichalcogenides (TMDs) hosting charge density waves (CDWs) provide a fertile playground for investigating symmetry-breaking phenomena \cite{wilson}. Among them, 1$T$-TiSe$_2$ is particularly intriguing due to its $2 \times 2 \times 2$ periodic lattice distortion (PLD) which emerges below $T_C \simeq 200$ K \cite{disalvo,holy,suzuki,chen,canadell,mauri}. While the CDW transition is widely accepted to be driven by a cooperative exciton-phonon mechanism \cite{cerc,monney,vWezel,kidd,kaneko, hedayat}, the symmetry properties of the ordered phase - specifically the emergence of chirality - 
sparks intense debate \cite{kim,chiC2A, chiC2B,chiC2C,mazin,chiral3,flu2}. Recent experiments using scanning tunneling microscopy (STM) \cite{chiral1,chiral2} and circular photogalvanic effect \cite{gedik} hinted at a chiral nature of the CDW state, while standard crystallographic analysis identified the ground state as the centrosymmetric space group $P\overline{3}c1$, which is achiral \cite{disalvo}. To resolve this paradox, theoretical models proposed that chirality arises from a specific phase relationship between the three CDW $q$-vectors (e.g., a $\pm2\pi/3$ mutual phase shift), leading to a "chiral hexatic" state \cite{jvw1,jvw2,nie}. Although phenomenologically appealing, the energetic stability of such phase-shifted configurations within the stiff potential landscape of the CDW was not rigorously verified. Furthermore, a distinction must be made between "spontaneous" chirality observed in surface-sensitive probes and macroscopic chiral order. Since chirality is strictly a 3D property, its detection via STM implies a specific coupling between the surface layer and the bulk stacking. This raises a fundamental question: does 1$T$-TiSe$_2$ possess an intrinsic chiral ground state, or is chirality a metastable feature stabilized by external fields or stacking faults? In this work, the origin of chirality in 1$T$-TiSe$_2$ is reassessed combining first-principles calculations with a phenomenological Landau theory. The previously proposed phase-shifted states are energetically unfavorable compared to amplitude variations, and the relevant degree of freedom is instead the CDW phase stacking order of the CDW layers. Density functional theory (DFT) results show that, while the ground state is indeed the achiral $P\overline{3}c1$ structure, a chiral phase with $C2$ symmetry lies only a few $\mu$eV per formula unit above it. A "dual-basis" description of the CDW recognizes that charge accumulation is bond-modulated while lattice displacement is atom-centered transverse. This geometric insight reveals how circularly polarized light can couple to the order parameter, lifting the degeneracy between chiral domains and inducing a macroscopic gyrotropic state. The model provides a unified explanation for the statistical appearance of chiral domains in dark cooling experiments and the controlled induction of chirality under optical drive.

\begin{figure}[t]
\includegraphics[width=85mm]{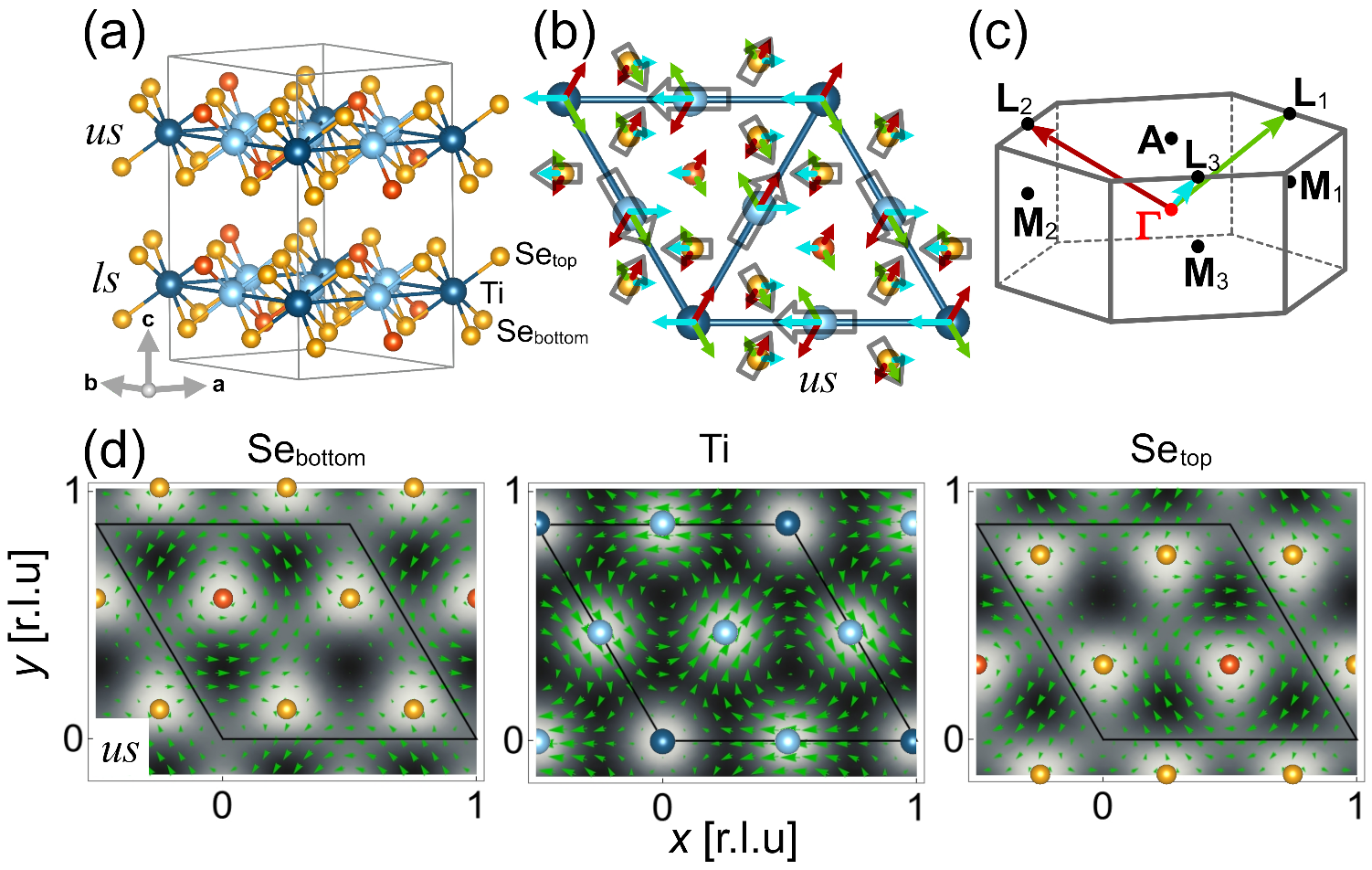}
\caption{
(a) $2\times 2\times 2$ CDW supercell of 1$T$-TiSe$_2$. Ti atoms are represented in blue, Se atoms in orange. Darker color labels atoms unaffected by the reconstruction. (b) The PLD in the upper slab, {\it us} (open arrows). The full PLD is the sum of three vector fields, each represented by a different color. (c) First BZ of the undistorted lattice with relevant symmetry points and propagation vectors ${\bf L}_j$. (d) Local PLD (green vector field) according to Eq.~\ref{eq2}: the left/middle/right column refers to the Se$_{\text{bottom}}$/Ti/Se$_{\text{top}}$ layer of the $us$ slab. Gray color scale highlights the charge density of the undistorted lattice (see Supplementary Note 1 for derivation details \cite{SM}).} \label{fig1}
\end{figure}

{\it The dual-basis order parameter} $-$
The $2\times2\times2$ lattice reconstruction of $1T$-TiSe$_2$ below $T_C \simeq 200$~K comprises two distinct Se-Ti-Se slabs \cite{disalvo}, labeled $ls$ (lower slab) and $us$ (upper slab) in Fig.~\ref{fig1}a.
Within each slab, titanium atoms (blue) form a covalently bonded sheet sandwiched between two hexagonal selenium layers (orange), shaping the octahedral coordination of Ti. Fig.~\ref{fig1}b illustrates the PLD pattern for the $us$ slab (open arrows); due to inversion symmetry, the displacement vectors in the neighboring $ls$ slabs are identical in magnitude, but opposite in sign (space group $P\bar{3}c1$). These displacements are strictly in-plane, with magnitudes $\delta_{\text{Ti}} \sim 3 \delta_{\text{Se}} \sim 0.085$~\AA, and arise from the superposition of the three vector fields (colored arrows) sketched in Fig.~\ref{fig1}b \cite{disalvo}. Due to the underlying trigonal symmetry, the complex order parameter $\Psi(\mathbf{r})$ that describes the CDW phase transition is conventionally expanded as a sum of three components \cite{mcmillan}:
\begin{equation}
\Psi({\bf r})=\sum_{j=1}^3 \Psi_j e^{i ({\bf q}_j \cdot {\bf r}+\phi_j)} \label{eq1}
\end{equation}
where $\Psi_j$ represent the real amplitudes, $\mathbf{q}_j$ the propagation vectors and $\phi_j$ the respective phases. It is crucial to distinguish the physical interpretation of this order parameter depending on the property being described. In the specific case of the lattice distortion in $1T$-TiSe$_2$, Eq.~\ref{eq1} defines the PLD when the propagation vectors correspond to the three symmetry-equivalent $L_j$ points at the boundary of the Brillouin zone (BZ) (Fig.~\ref{fig1}c) \cite{disalvo,mauri,canadell,suzuki,kaneko}.
According to standard supercell conventions, the parent-zone $L(M)$ points are folded back to the $\Gamma(A)$ point of the reduced $2\times2\times2$ zone.
The observed atomic displacements $\delta \mathbf{u}$ correspond to the eigenvectors of the $L_1^-$ soft transverse optical phonon mode of the undistorted phase. Consequently, the theoretically predicted distortion is strictly transverse to these wavevectors \cite{mauri,canadell,kaneko}, and naturally follows the relation $\delta \mathbf{u} \propto  {\bf \hat{z}} \times {\bf L}_j$ (where ${\bf \hat{z}}$ is the slab normal) with the specific displacement ratio $\delta_{Ti} \sim 3\delta_{Se}$. Equivalently, this can be expressed using the in-plane projections $\mathbf{M}_j$ as $\delta \mathbf{u} \propto {\bf \hat{z}} \times \mathbf{M}_j$, leading to the compact form:
\begin{equation}
\delta{\bf u(r}) \propto {\bf \hat{z}}\times \R[\nabla \Psi({\bf r})] = -\sum_{j=1}^3 \Psi_j {\bf \hat{z}} \times {\bf M}_j \sin({\bf L}_j \cdot {\bf r}+\phi_j) \label{eq2}
\end{equation}
By selecting uniform amplitudes $\Psi_j = \Psi_0$ and vanishing phases $\phi_j = 0$, this expression perfectly reproduces the experimental displacement pattern consistent with the Wyckoff positions of the $P\bar{3}c1$ space group \cite{wyckoff} (see Supplementary Note 1-2 \cite{SM}). This vector field is visualized in Fig.~\ref{fig1}d (green arrows), where the left/middle/right panels correspond to Se$_{\text{bottom}}$/Ti/Se$_{\text{top}}$ layer of the $us$ slab, respectively.

\begin{figure}[htb]
\includegraphics[width=85mm]{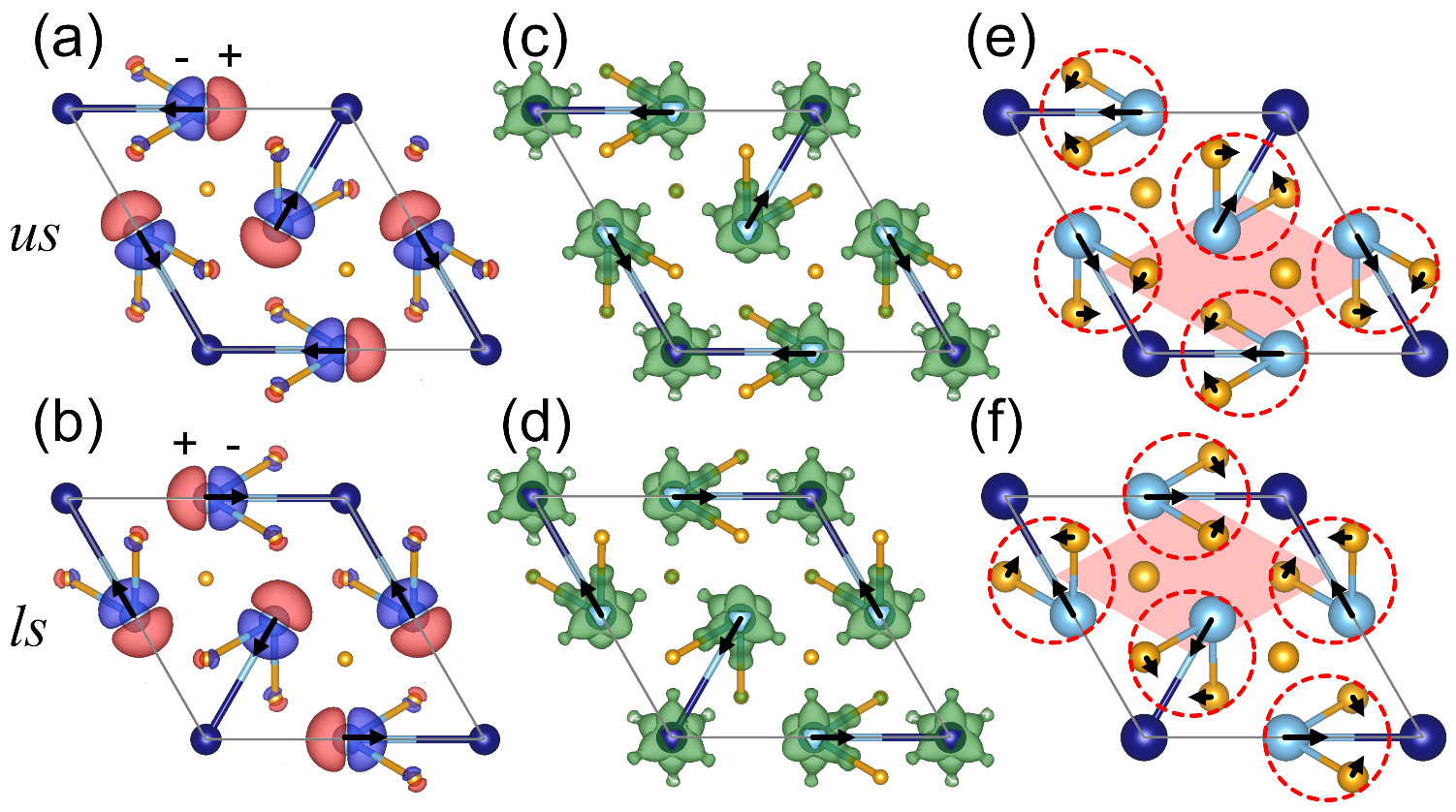}
\caption{Top/bottom panels refer to the $us/ls$ slab. (a)-(b) Difference between the charge distribution of the CDW phase and that of the undistorted lattice. The blue (negative) and red (positive) lobes are a consequence of the atomic displacements in the charge-ordered phase. (c)-(d) Isosurface charge density (green, shaded shapes) after removing the superposition of the atomic ones, showing the bonding/anti-bonding character. (e)-(f) TiSe$_2$ clustering, highlighted by red dashed circles.}
\label{fig2}\end{figure}

\begin{figure}[htb]
\includegraphics[width=85mm]{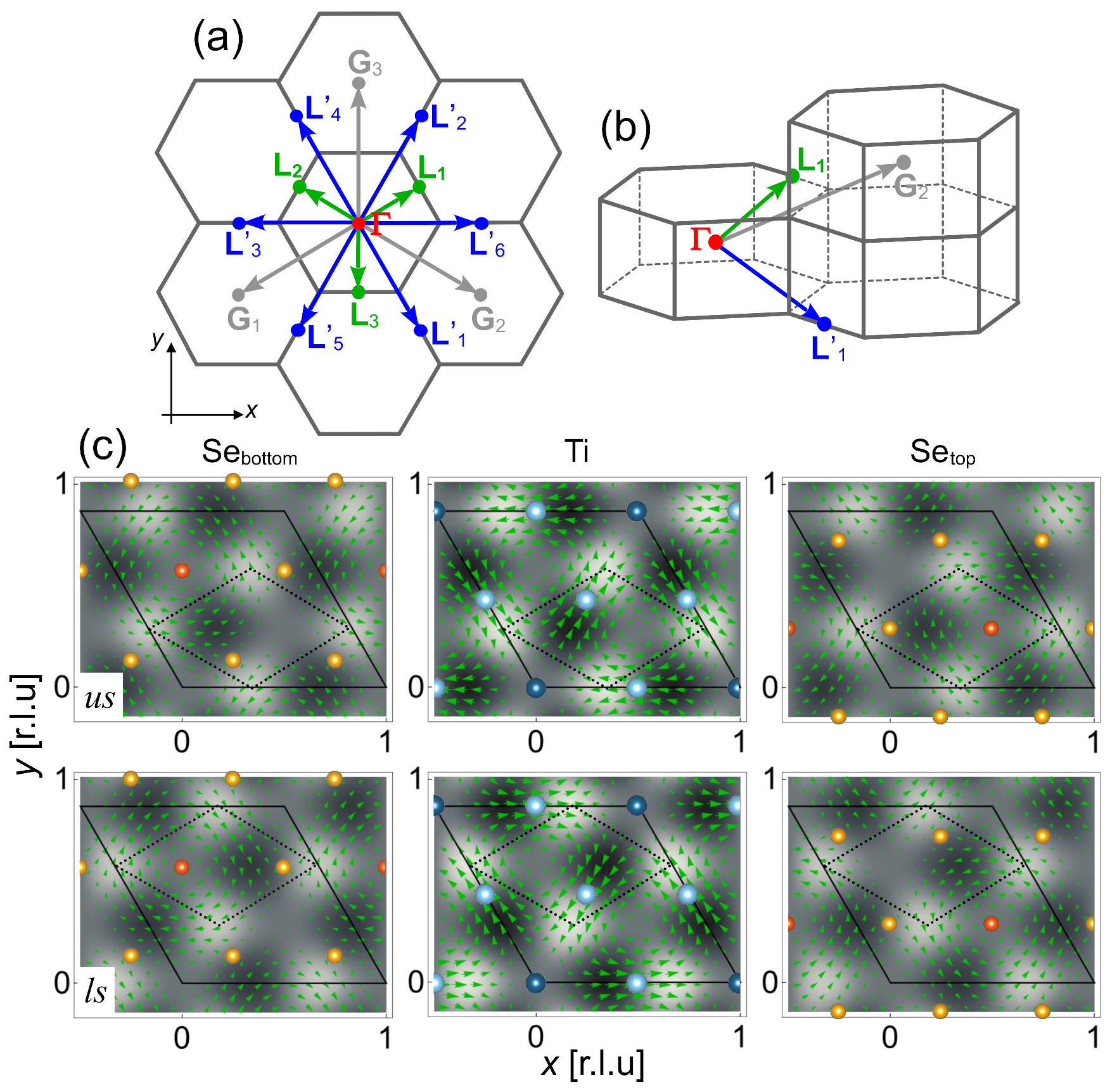}
\caption{(a) Propagation vectors (colored arrows) used in Eqs.~\ref{eq2}-\ref{eq3}, projected on the $xy$ reciprocal plane, and (b) 3D rendering of representative ones within undistorted BZs (gray). (c) Local PLD (green vector field) and the corresponding charge density (gray color scale) according to Eqs.~\ref{eq2}-\ref{eq3}, respectively. The left/middle/right column refers to the Se$_{\text{bottom}}$/Ti/Se$_{\text{top}}$ layer; the upper/lower row refers to the $us/ls$ slab. Black solid and dotted sub-shells highlight the corresponding PLD and charge periodicities.}
\label{fig3}\end{figure}

While Eq.~\ref{eq2} accurately captures the lattice response, the associated charge redistribution requires a more nuanced description. The background gray scale in Fig.~\ref{fig1}d approximates the charge density $\rho(\mathbf{r})$ of the undistorted lattice, which follows the periodicity of the reciprocal lattice vectors $\mathbf{G}_h$ (see Supplementary Note 1 for derivation details \cite{SM}). However, to understand the induced charge distribution, density functional theory (DFT) is deployed. Using the PBE-GGA functional with Grimme-D3 van der Waals correction, the fully relaxed CDW structure modeled with the $P\overline{3}c1$ space group provides the best agreement with the experimental lattice parameters and distortion pattern (see Supplementary Note 2 \cite{SM}). The DFT results, displayed in Fig.~\ref{fig2}, reveal a striking feature: the charge redistribution is not localized at the atomic sites, but is instead distributed along the length of the chemical bonds (i.e., it is bond-modulated).
The charge density difference between undistorted and reconstructed lattice structures (Figs.~\ref{fig2}a-b) shows that charge accumulation (blue lobes) occur primarily along the Ti-Se bonds. Similarly, the net charge density after removal of the superimposed atomic ones (green, shaded shapes in Figs.~\ref{fig2}c-d) clearly shows bonding/anti-bonding characters, matching the blue/red lobes of the previous panels and occurring at the sites of TiSe$_2$ clustering. The latter are highlighted by red dashed circles in Figs.~\ref{fig2}e-f, each involving a single TiSe$_2$ formula unit (f.u.). This implies that the CDW-induced electronic order is bond-modulated and cannot be described by the $\mathbf{L}_j$ periodicity (defined in Fig.~\ref{fig1}c) that acts on alternating atomic sites. It rather follows the pattern outlined by the red-shaded hexagonal sub-cells in  Figs.~\ref{fig2}e-f. An intuitive way to deduce the electronic distribution $\rho_C(\mathbf{r})$ in the charge-ordered phase is to realize that it spawn from the atomic charge density $\rho(\mathbf{r})$ in the undistorted structure being displaced by the lattice deformation: taking the leading-order approximation in the small-displacement limit, it is $\rho_C(\mathbf{r}) \sim \rho({\bf r}+\delta{\bf u})-\rho({\bf r}) \sim \delta{\bf u(r}) \cdot \nabla \rho({\bf r})$. A detailed calculation (see Supplementary Note 1 \cite{SM}) leads to:
\begin{equation}
\rho_C({\bf r})=\delta{\bf u(r}) \cdot \nabla \rho({\bf r}) \propto \sum_{g=1}^3 \Psi_g [\cos({\bf L}'_g \cdot {\bf r}) - \cos({\bf L}'_{g+3} \cdot {\bf r}) ]\label{eq3}
\end{equation}
Since $\rho({\bf r})$ and $\delta{\bf u(r})$ have different periodicities (${\bf G}_h$ and ${\bf L}_j$, respectively), the spatial modulation of $\rho_C({\bf r})$ combines ${\bf G}_h$ and ${\bf L}_j$ in the resulting propagation vectors ${\bf L}'_g$ represented in Figs.~\ref{fig3}a-b. They all have equal lengths and fold the $\Gamma$(A) point onto the L(M) points of the undistorted BZ (like ${\bf L}_j$), but involve neighbouring zones.
Most importantly, they outline the red-shaded sub-cells in Figs.~\ref{fig2}e-f. Eqs.~\ref{eq2}-\ref{eq3} are graphically rendered in Fig.~\ref{fig3}c with the same panel sequence used in Fig.~\ref{fig1}d. The top/bottom row refers to the $us$/$ls$ slabs, respectively. While the periodicity of the PLD (green vector field, Eq.~\ref{eq2}) is defined by the black, solid unit cell, the corresponding charge density (gray color scale, Eq.~\ref{eq3}) is carved within the dotted sub-cells, in perfect agreement with the DFT results (Fig.~\ref{fig2}). Although, both PLD and CDW derive from the same order parameter (Eq.~\ref{eq1}), the fact that the charge density is physically rooted in the bonding orbitals (${\bf L}'_g$ basis), while the lattice reconstruction manifests geometrically as a transverse mode (${\bf L}_j$ basis) has profound consequences on the optically-driven properties of 1$T$-TiSe$_2$, as clarified in the next sections.

{\it Landau energy landscape and CDW phase stacking order} $-$
To determine the true ground state and assess the stability of potential chiral configurations, a systematic exploration of the free energy landscape combining DFT and the Landau theory of phase transitions is performed, focusing on the interplay between the amplitude ($\Psi_j$), phase ($\phi_j$) and stacking order of the CDW. A prevailing hypothesis for the origin of chirality in 1$T$-TiSe$_2$ relies on a mutual phase shift between the three charge density components (e.g., $\Delta \phi=\pm 2\pi/3$) \cite{chiral1,gedik,jvw1,jvw2}. This hypothesis is rigorously tested by calculating the total energy as a function of the relative phase $\phi_j$ while keeping the amplitude fixed at the equilibrium value. As detailed in Supplementary Note 3 \cite{SM}, the analysis reveals a remarkably stiff potential well: any deviation from the in-phase condition ($\phi_j=0$) incurs a prohibitive energy cost as compared to amplitude deviations. This finding effectively rules out phase fluctuations as the primary driver of chirality, compelling the search for alternative symmetry-breaking mechanisms. Having excluded phase shifts, the attention is turned to the stacking degrees of freedom, where the Landau model of phase transitions proves decisive.

\begin{figure*}[htb]
\includegraphics[width=165mm]{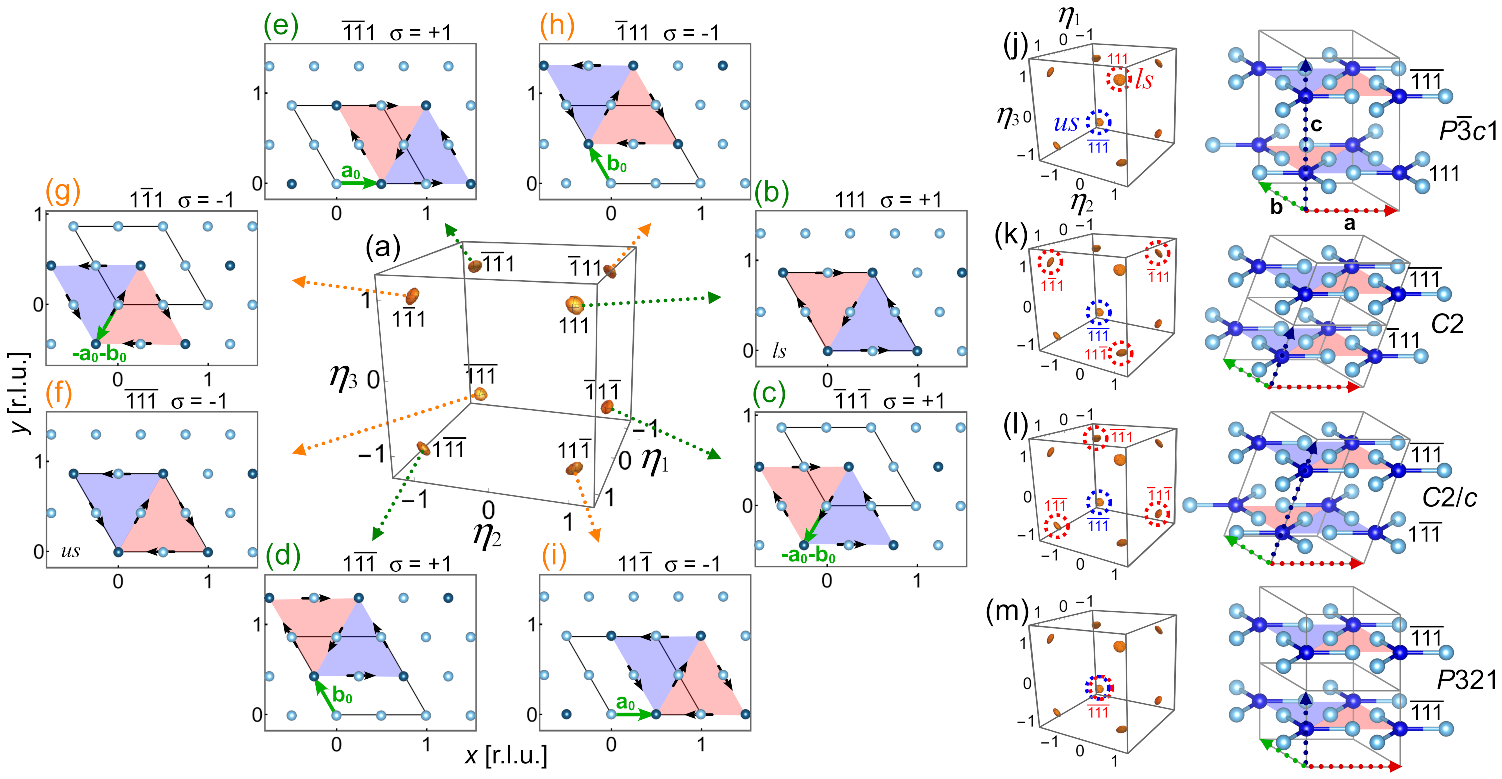}
\caption{(a) The eight equivalent absolute minima of the free energy (Eq.~\ref{eq9}) for a single 1$T$-TiSe$_2$ slab. Each minimum is labeled by its fraction coordinates. (b)-(i) PLD patterns, computed via Eq.~\ref{eq2} (the {\it ls} slab is used as benchmark), corresponding to each free energy minimum in panel (a); only Ti atoms are represented for clarity; ${\bf a}_0$ and ${\bf b}_0$ are lattice vectors of the undistorted structure; blue- and red-shaded triangles highlight the PLD parity. (j)-(m) The four non-equivalent double-slab combinations; the configuration of the $us$ slab is assumed $\overline{111}$ and the blue circles [left panels] mark the corresponding free energy minimum. Different $ls$ slabs give rise to specific CDW phase stacking and crystal symmetries [right panels]. The red circles in (k)-(l) [left panels] mark $ls$ slabs that lead to equivalent crystals under $C_3$ symmetry operation.}\label{fig4}
\end{figure*}

The free energy of a three-component order parameter with trigonal symmetry (Eq.~\ref{eq1}) is \cite{cowley} (see Supplementary Note 4 \cite{SM}):
\begin{widetext}
\begin{equation}
G(T)-G_1 = \Psi_0^4 \left[- \frac{3 \mu+\nu}{2}\left(1-\frac{T}{T_C}\right) (\eta_1^2+\eta_2^2+\eta_3^2)+\frac{\mu}{4} (\eta_1^2+\eta_2^2+\eta_3^2)^2+\frac{\nu}{4} (\eta_1^4+\eta_2^4+\eta_3^4)\right] \label{eq9}
\end{equation}
\end{widetext}
Here, the coefficients $\mu$ and $\nu$ (both positive) are temperature-independent, and the amplitudes $\Psi_j = \Psi_0 \eta_j$ are expressed using dimensionless fractional coordinates $\eta_j$. Below the critical temperature $T_C$, the free energy landscape exhibits eight equivalent global minima located at $\tilde{\eta}_j = \pm \sqrt{1-T/T_C}$. At $T=0$ K, Eq.~\ref{eq9} describes the energetics of a single, isolated $1T$-TiSe$_2$ slab in its CDW ground state. The constant $G_1$ is chosen so that $G(T=0\text{ K}) = 0$ at any minimum (see Supplementary Note 4A \cite{SM}). By mapping the fractional coordinates to a Cartesian frame (Fig.~\ref{fig4}a), these eight minima are identified by the orange disks and labeled by their coordinates triplet (e.g., the state $\eta_1=\eta_2=1, \eta_3=-1$ is denoted as '$11\bar{1}$').
The corresponding periodic lattice distortion patterns, calculated via Eq.~\ref{eq2}, are displayed in Figs.~\ref{fig4}b-i. These configurations differ in two key aspects: (i) a relative rigid shift corresponding to a lattice vector of the undistorted cell [$\pm \mathbf{a}_0, \pm \mathbf{b}_0$ or $\pm(\mathbf{a}_0+\mathbf{b}_0)$]; and (ii) the parity of the PLD, defined by the product $\sigma = \eta_1 \eta_2 \eta_3 = \pm 1$. This parity is visually distinguished by the orientation of the blue and red shaded triangles in Fig.~\ref{fig4} (configurations with $\sigma=+1$ and $\sigma=-1$ are shown in panels \ref{fig4}b-e and \ref{fig4}f-i, respectively).

Consequently, when stacking two slabs to form the bulk crystal, only four unique, non-equivalent arrangements emerge (Fig.~\ref{fig4}j-m). Assuming a fixed configuration for the upper slab ($us$) - for instance, the $\bar{1}\bar{1}\bar{1}$ state with $\sigma_{us}=-1$ - the lower slab ($ls$) can adopt one of four distinct relationships:
\begin{itemize}[leftmargin=*]
\item[-] Fig.~\ref{fig4}j ($P\bar{3}c1$): No relative rigid shift and opposite PLD parity ($\sigma_{ls}=+1$). This yields the centrosymmetric $P\bar{3}c1$ space group, corresponding to the standard crystallographic description of the CDW phase.
\item[-] Fig.~\ref{fig4}k ($C2$): A relative rigid shift combined with the same PLD parity ($\sigma_{ls}=-1$). This configuration belongs to the monoclinic $C2$ space group, which lacks both inversion and mirror symmetries.
\item[-] Fig.~\ref{fig4}l ($C2/c$): A relative rigid shift with opposite PLD parity ($\sigma_{ls}=+1$), resulting in the centrosymmetric $C2/c$ space group.
\item[-] Fig.~\ref{fig4}m ($P321$): No relative rigid shift with the same PLD parity ($\sigma_{ls}=-1$), corresponding to the non-centrosymmetric $P321$ space group.
\end{itemize}
Notably, both the $C2$ and $P321$ phases belong to the Sohncke space groups, which contain only proper rotation operations and therefore support chirality \cite{cfelser}. Unlike the isolated slabs, these four CDW phase stacking arrangements are not energetically degenerate due to interlayer coupling. DFT calculations reveal that the ground state $P\bar{3}c1$ (Fig.~\ref{fig4}j) and the chiral $C2$ phase (Fig.~\ref{fig4}k) are nearly isoenergetic, separated by only $4.7~\mu\text{eV/f.u.}$ In contrast, the $C2/c$ and $P321$ configurations lie significantly higher in energy ($26~\mu\text{eV/f.u.}$ and $59~\mu\text{eV/f.u.}$, respectively). This energetic proximity suggests that the chiral $C2$ structure is highly accessible when the CDW sets in. Furthermore, statistical considerations favor the chiral phase: the $C2$ stacking can be realized via three equivalent minima (labeled $1\bar{1}1, \bar{1}11, 11\bar{1}$ in  Fig.~\ref{fig4}k), making it three times more probable than the unique $P\bar{3}c1$ stacking.

\begin{figure}[t]
\includegraphics[width=85mm]{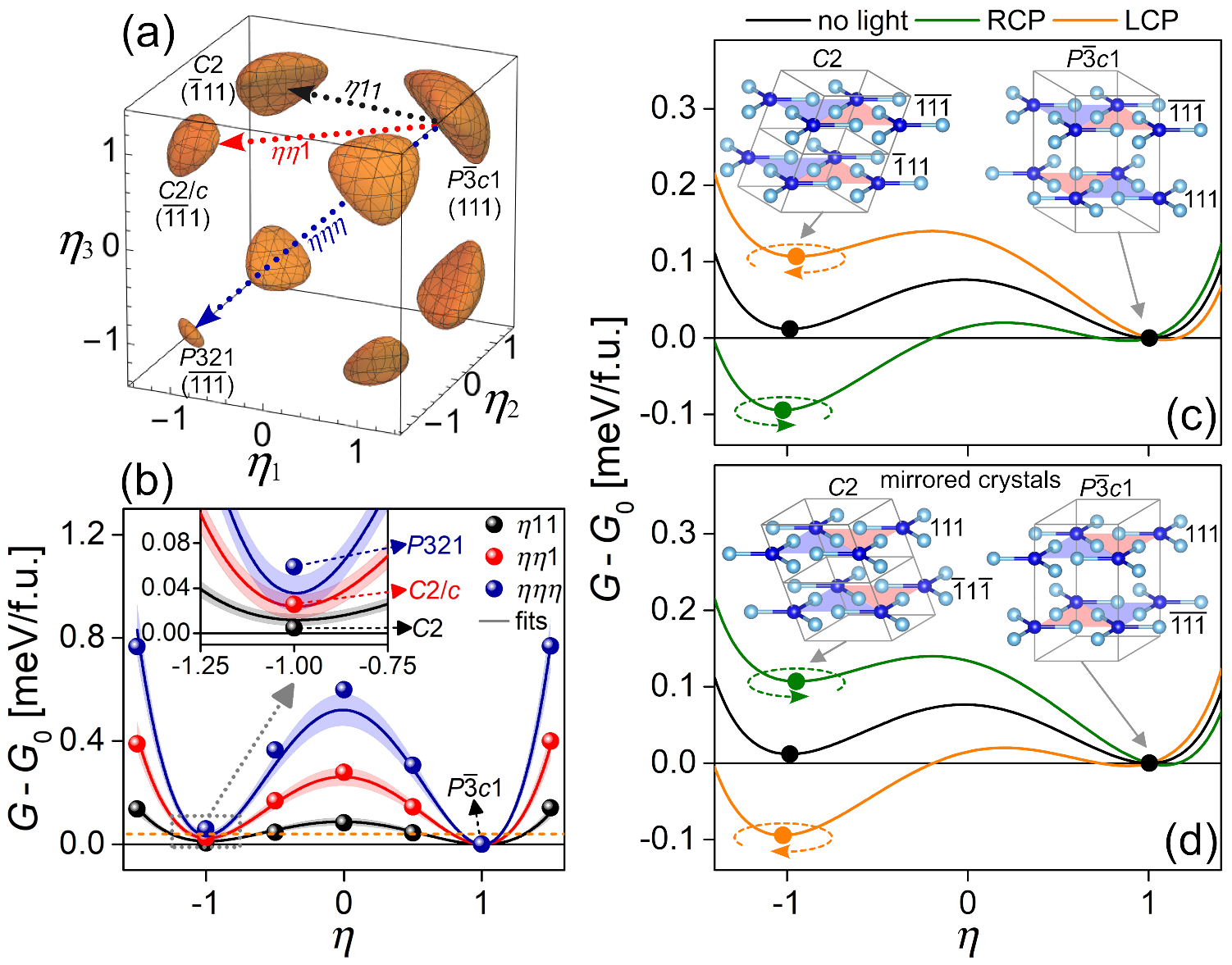}
\caption{(a) Constant free energy contour plot including the interlayer coupling (Eqs.~\ref{eq9}-\ref{eq10}). Coloured arrows indicate the paths along which the $ls$ slabs is modified to compute the free energy barriers separating non-equivalent stacking ($us$ slab is fixed in the $\overline{111}$ configuration). (b) Total free energy along the paths described in panel (a); dots refer to DFT calculations, solid lines are fits of the DFT results with Eqs.~\ref{eq9}-\ref{eq10} (the shaded areas are intervals of confidence). The orange dashed line marks the energy of the contour plot in panel (a). The inset highlights the effect of interslab coupling (Eq.~\ref{eq10}). (c) Free energy barrier between $P\bar{3}c1$ and $C2$ structures at $T=0$~K, including the chiral induction term (Eq.~\ref{eq12} - numerical values $|\mathbf{E}| = 1$ and $\delta/\Psi_0^3 = 0.05$ are used), in the absence of illumination (black curve, same as panel (b)), and under RCP (green curve) or LCP light (orange curve). (d) Same as panel (c) but applied to the mirror-symmetric crystal structures; the effect of RCP and LCP light swaps, establishing enantiomeric behaviour of the $C2$ stacking.}
\label{fig5}
\end{figure}

It is instructive to quantify the free energy barriers separating the four non-equivalent stackings. Referring to Fig.~\ref{fig5}a, gradually changing $\eta_1$ while keeping $\eta_2=\eta_3=1$ turns the $P\overline{3}c1$ structure into $C2$. This path is labeled $\eta 11$ (black-dotted arrow). Similarly, varying two coordinates ($\eta\eta 1$, red-dotted arrow) or all three simultaneously ($\eta\eta\eta$, blue-dotted arrow) turn the $P\overline{3}c1$ stacking into $C2/c$ or $P321$, respectively. The PLD of each intermediate configuration is obtained via Eq.~\ref{eq2}, its total energy is computed by DFT and plotted as a function of the common fractional coordinate $\eta$, providing the energy landscape of the different transitions shown in Fig.~\ref{fig5}b (symbols). The energy scale is referred to the energy $G_0$ of the $P\overline{3}c1$ ground state, as obtained by DFT.
To account for interlayer coupling, the following term is added to the free energy (see Supplementary Note 4B \cite{SM}):
\begin{equation}
G_c -G_2= \gamma \Psi_0^2 (\eta_1,\eta_2,\eta_3)\cdot (\zeta_1,\zeta_2,\zeta_3) \label{eq10}
\end{equation}
Here, $\eta_j$ and $\zeta_j$ are the fractional coordinates of the {\it ls} and {\it us} slabs, while $\gamma>0$ is a phenomenological constant. The constant $G_2$ is chosen so that $G_c = 0$ when $\eta_j=1$ and $\zeta_j=-1$ ($j=1,2,3$), that is the $P\overline{3}c1$ double-slab ground state (see Supplemental Note 4B \cite{SM}). Eq.~\ref{eq10} accurately captures the energy hierarchy of the different stackings (Fig.~\ref{fig5}b, inset). The solid, coloured lines in panel~\ref{fig5}b fit of the DFT data according to Eqs.~\ref{eq9}-\ref{eq10} with a unique set of parameters $\mu$, $\nu$ and $\gamma$ (see Supplementary Note 4D \cite{SM}), establishing that the free energy barrier between $P\overline{3}c1$ and $C2$ structures is the narrowest. This approach provides the link between {\it ab-initio} DFT calculations and the phenomenological Landau model, disclosing the full energy landscape of the CDW phase, but most importantly it lays the foundations to clarify the highly debated aspect of charge ordering in $1T$-TiSe$_2$: its chiral nature.

{\it Chiral induction} $-$
The identification of the nearly degenerate $C2$ phase provides the necessary substrate for chirality, but it leaves one question unanswered: how does an external field, specifically circularly polarized light, select a preferred handedness from the random thermal distribution? The answer lies in the unique "dual-basis" nature of the CDW order parameter. As previously established, the charge density $\rho_C(\mathbf{r})$ is bond-modulated (described by $\mathbf{L}'_g$ basis), while the lattice distortion $\delta\mathbf{u}(\mathbf{r})$ is transverse atom-centered (described by $\mathbf{L}_j$ basis). This distinction enables a specific optical coupling pathway that is forbidden in a single-basis description. By expanding the free energy to include the electric field of light, $\mathbf{E}$, the lowest-order chiral coupling term $G_\chi$ is derived. It arises from the interaction between the bond-modulated charge density and the transverse lattice distortion, taking the form (see Supplementary Note 4C \cite{SM}):
\begin{equation}
G_\chi = \delta \Psi_0^3 \left[ (\mathbf{E} \times \mathbf{E}^*) \cdot {\bf \hat{z}} \right] (\sigma_{ls}  + \sigma_{us}) \label{eq12}
\end{equation}
Here, $\delta$ is a positive phenomenological constant, the term $(\mathbf{E} \times \mathbf{E}^*) \cdot {\bf \hat{z}}$ is directly proportional to the helicity of the light (positive for right circular - RCP, negative for left circular polarization - LCP) and the factor $(\sigma_{ls}  + \sigma_{us})$ is the combined PLD parity of the double-slab. This expression dictates strict selection rules for the induction of chirality. Fig.~\ref{fig5}c shows the free energy barrier separating the $P\bar{3}c1$ and $C2$ structures ($\eta 1 1$ path) at $T=0$~K in the absence of illumination (black curve, same as Fig.~\ref{fig5}b), and under RCP (green curve) or LCP light (orange curve). In the $P\bar{3}c1$ ground state, the two slabs have opposite parities ($\sigma_{ls} = -\sigma_{us}$), causing their contributions to cancel exactly ($G_\chi = 0$). Consequently, the bulk ground state is optically inert to chiral induction. In contrast, for the $C2$ stacking the parity alternation is frustrated ($\sigma_{ls} = \sigma_{us}$), leading to a constructive addition of the chiral term ($G_\chi \neq 0$). Under circularly polarized illumination, this term acts as a symmetry-breaking field that lowers the free energy for one light helicity relative to the other. If the mirror-symmetric crystal structures are considered (Fig.~\ref{fig5}d), the roles of RCP and LCP light swap, implying that light helicity can boost one $C2$ domain at expenses of its enantiomer. As the system cools through $T_C$ under illumination, this energetic bias tilts the effective potential, favoring the nucleation and growth of "light-selected" domains. The result is a macroscopic, single-domain chiral state that persists even after the light is removed, consistent with the observations of light-induced gyrotropy \cite{gedik}. It is relevant to point out that at any finite temperature below $T_C$, the fractional coordinates scale as $\sqrt{1-T/T_C}$. Accordingly, the interslab coupling term $G_c \propto (1-T/T_C)$ (Eq.~\ref{eq10}) is vanishing small for $T\lesssim T_C$, making the $P\overline{3}c1$ and $C2$ structures strongly degenerate. In the absence of helical light, thermal fluctuations will populate the $C2$ basins. Since these form with random handedness ($C2$ enantiomers being degenerate without light), the macroscopic sample will be a racemic mixture of chiral domains (see Supplementary Note 5 \cite{SM}), appearing "achiral" in bulk average (e.g., zero net optical activity), but "chiral" to local probes (STM). This provides a natural explanation for "spontaneous" chirality .

In summary, a new microscopic framework for understanding the chiral CDW phase in 1$T$-TiSe$_2$ is established. First-principles calculations rule out the mutual phase-shift model, showing that the energy penalty for de-phasing the order parameters is too high to support a stable ground state. Instead, the competition between the centrosymmetric $P\overline{3}c1$ ground state and the chiral $C2$ stacking is the central mechanism. The near-degeneracy of these states implies that, upon cooling in the absence of external fields, the system is dominated by entropic fluctuations, resulting in a racemic mixture of chiral $C2$ domains. This successfully explains the "spontaneous" chirality frequently observed in local STM measurements without violating the global symmetry constraints of the bulk crystal. Crucially, the optical selection rules governing this transition is derived by introducing a dual-basis description: the CDW cannot be fully described by a single set of harmonic vectors; rather, the charge density is physically located on Ti-Se bonds, while the lattice distortion is transverse and atom-centered. This distinction is not merely a mathematical formality, but it is physical rooted in the nature of the 1$T$-TiSe$_2$ CDW phase, ultimately permitting a non-zero coupling between the order parameter and the helicity of light. The concept of a dual-basis decoupling - where electronic charge redistribution and structural deformations exhibit distinct spatial periodicities - may be a more general feature. Similar symmetry-breaking mechanisms could be at play in other charge-ordered transition metal dichalcogenides leading to the formation of  bonding/antibonding orbitals coupled to transverse distortions. These findings suggest that 1$T$-TiSe$_2$ belongs to a class of materials where hidden degrees of freedom - in this case,  the CDW phase stacking order - can be manipulated by light to control global symmetry, opening the door to ultrafast optical switching of chirality in TMDs, with potential applications in chiroptical memory and switchable non-linear optical devices.


\end{document}